\documentclass[aps,amsmath,amssymb,preprintnumbers,groupedaddress,twocolumn,floatfix,superscriptaddress,nofootinbib]{revtex4-2}
\usepackage{amsmath}
\usepackage{amsfonts}
\usepackage{amssymb}
\usepackage{stmaryrd}
\usepackage{empheq}
\usepackage[arrow,matrix,curve]{xy}
\usepackage{graphicx} 
\usepackage{pstricks}
\usepackage[hidelinks]{hyperref}
\newcommand{\nc}{\newcommand}
\nc{\beq}{\begin{equation}}
\nc{\eeq}{\end{equation}}
\nc{\bea}{\begin{eqnarray}}
\nc{\eea}{\end{eqnarray}}

\newcommand{\eq}[1]{\begin{equation}
	\begin{split} #1 \end{split}
	\end{equation}}

\begin{document}


\preprint{MPP-2021-110}

\title{Swampland Conjectures for an Almost Topological Gravity Theory}
	
\author{Rafael \'Alvarez-Garc\'ia}
\affiliation{II. Institut f\"ur Theoretische Physik, Universit\"at Hamburg,
		Luruper Chaussee 149, 22607 Hamburg, Germany}
\author{Ralph Blumenhagen}
\affiliation{Max-Planck-Institut f\"ur Physik, F\"ohringer Ring
		6, 80805 M\"unchen, Germany}
\author{Christian Knei\ss l}
\affiliation{Max-Planck-Institut f\"ur Physik, F\"ohringer Ring
		6, 80805 M\"unchen, Germany}
\affiliation{Ludwig-Maximilians-Universit\"at M\"unchen, Fakult\"at f\"ur Physik, Theresienstr. 37, 80333 M\"unchen, Germany}
\author{Andriana Makridou}
\affiliation{Max-Planck-Institut f\"ur Physik, F\"ohringer Ring
		6, 80805 M\"unchen, Germany}
\author{Lorenz Schlechter}
\affiliation{Max-Planck-Institut f\"ur Physik, F\"ohringer Ring
		6, 80805 M\"unchen, Germany}


\begin{abstract}
We analyze $AdS$ and $dS$  swampland conjectures
in a three-dimensional higher spin  theory with self-interacting matter, which  
contains conformal gravity and  is almost topological. A theory of
a similar type  was proposed as the effective theory in the high energy phase
of non-critical M-theory in 3D. With
some details differing from the usual string theory story,
it is found that the resulting effective theory, namely
topologically massive gravity,  fits  well into the web of the
proposed swampland conjectures. 
Supporting a recent proposal, in particular we find that this 3D theory gives rise to a quantum break
time that scales like $t_Q\sim H^{-1}$.
\end{abstract}


\maketitle


\section{Introduction}
\label{sec:intro}

The swampland program~\cite{Vafa:2005ui} implies a paradigm shift  in the field of string
phenomenology. Instead of analyzing the string landscape
for certain well understood classes of string compactifications,
one tries to mark the boundary between the landscape and the swampland.
This is quantified in the form of a number of swampland conjectures,
which are supported by 
concrete string models and/or
by more general quantum gravity arguments (see \cite{Palti:2019pca, vanBeest:2021lhn,Grana:2021zvf} for recent reviews).
In this endeavor, our intuition is led by experience from critical string
theory or M-theory and their low-energy effective actions. 
As such, we have a universal gravity sector that is governed
by an Einstein-Hilbert action that  upon
quantization leads to gravitons interacting with a strength set by the
Planck mass. 

The purpose of this letter is to consider effective gravity theories 
that are of a different kind but have still appeared in the string
theory literature, hence having a good chance to be in the string
landscape. We want to challenge such theories with some of the swampland
conjectures,  in particular the $AdS$ and $dS$  swampland conjectures,
as well as the proposal of quantum breaking.

One instance where such gravity  theories of a different kind are expected to
appear  is in the high energy or high temperature regime of string theory.
Unfortunately, for temperatures beyond the Hagedorn transition these
theories are not completely understood, but there are indications, like the quadratic scaling of the
free-energy with temperature \cite{Atick:1988si},  that the number of degrees of
freedom gets reduced.  This letter is also motivated by the recent
proposal \cite{Blumenhagen:2020doa} that
in the high temperature regime of string theory, the quantum break
time should scale like  $t_Q\sim H^{-1}$ irrespective of the number of
dimensions. Recall that, as shown in \cite{Dvali:2017eba}, 
for Einstein gravity  in $d$ space-time dimensions this
scaling is $t_Q\sim M_{\rm pl}^{d-2}/H^{d-1}$.

Clearly, for making progress  we need a concrete and well treatable starting ground. 
For our purposes, this is provided by the  non-critical M-theory
in 3D proposed by Ho\v{r}ava/Keeler
\cite{Horava:2005tt,Horava:2005wm,Horava:2007ds}  
which is exactly solvable in terms of a non-relativistic Fermi liquid
in 2 +1 dimensions. This is motivated by the known tachyon
condensation from ten-dimensional type 0 theories, which represent the
infinite temperature limit of type II theories, to the two-dimensional
type 0 theories \cite{Hellerman:2007fc}.  The non-critical M-theory
arises from the thermal version of this two-dimensional theory when
D0-branes are included \cite{Horava:2005tt,Horava:2005wm,Horava:2007ds}.
In the conformal limit this theory was suggested to be described
by an effective higher spin generalization of conformal gravity
equipped  with a higher spin matter field.
The ground state of non-critical M-theory corresponds 
to an $AdS_2\times S^1$ solution of this theory plus a massless fermion.
Without the matter field,  the theory would be topological
with no propagating degrees of freedom.
Note that such topological theories  were also invoked in the recent proposal
\cite{Agrawal:2020xek} of a topological phase in the early universe. 

This higher spin theory
describing the conformal limit of non-critical M-theory
is motivating our choice of a non-standard effective gravity theory.
The latter is obtained as a truncation to just conformal gravity,
where we will also include a conformally coupled, scalar matter
field\footnote{Note that the
truncation may in principle have qualitatively different properties from its higher
spin completion. In particular, the much bigger gauge invariance of
the higher spin theory may render some solutions pure
gauge. We thank the referee for raising this point.}.
This is expected to decouple in the strict conformal limit,
but is  still  present at large but finite energies. 
By also including a conformal self-interaction of the scalar field,
this truncated theory admits non-trivial $AdS_3$ and $dS_3$ solutions for 
a non-vanishing vacuum expectation value
of the scalar. The resulting effective theory
is known as topologically massive gravity
(TMG) \cite{Deser:1981wh}, carrying the wrong sign of the 
Einstein-Hilbert term.
In this letter we consider this almost topological theory as a toy model for the
high energy phase of string theory and 
analyze  whether it  does still  satisfy in particular those swampland conjectures dealing
with $dS$ and $AdS$ solutions.

\section{Non-critical M-theory in 3D}
\label{sec_two}

Three-dimensional non-critical M-theory was  proposed in
\cite{Horava:2005tt,Horava:2005wm} as the strong coupling limit
of non-critical type 0A theory in 2D.
In \cite{Horava:2005tt} a 
non-perturbative definition of this theory was formulated  as a
double-scaled non-relativistic Fermi liquid in $2+1$ dimensions.
This theory is closely related to the matrix model description
of 2D type 0A theory by summing over all D0-brane charges.
Moreover, it  has an infinite dimensional  symmetry described
by a bosonic higher-spin Lie algebra, called  ${\cal W}_0$ in
\cite{Horava:2005wm}. Important in the following is that 
the conformal algebra $SO(3,2)$ is a finite dimensional subalgebra of ${\cal W}_0$.
For more details we refer the reader to the original literature.

In \cite{Horava:2005wm} it was found that in  the 
high-temperature regime the non-perturbative partition function  
is closely related to the one of the topological A-model on the
resolved conifold. 
In the conformal limit ($\alpha' \rightarrow \infty$) it was argued 
\cite{Horava:2007ds} that the ground state of non-critical M-theory 
should be an $AdS_2 \times S^1$  space-time accompanied
by  a massless free fermion as the only propagating
degree of freedom.

As proposed in \cite{Horava:2007ds}, a higher spin theory admitting such a solution is given
by the higher spin Chern-Simons theory \cite{Pope:1989vj}
\eq{
\label{HCSaction}
                   S_{\rm HCS}= {1\over 4} \int_{M} {\rm
                     Tr}\left({\cal A}\wedge d{\cal A} +{2\over 3} {\cal A}\wedge
                   {\cal A}\wedge {\cal A}\right)
}
with the gauge field one-form ${\cal A}$  taking values in the
infinite-dimensional higher-spin Lie-algebra ${\cal W}_0$.
As shown in \cite{Shaynkman:2001ip},  matter fields 
can be coupled to this topological theory at least at the level
of the matter field equations of motion in a CS background field ${\cal A}$ via
\eq{
                   d|\Phi\rangle+{\cal A}\star |\Phi\rangle=0\,,
}
where $|\Phi\rangle$ is a higher spin matter field that contains
a scalar and a fermion as the lowest components.
In an $AdS_2\times S^1$ (or also $AdS_3$)  background with all higher
spin fields set to zero, all these equations  reduce  to two equations of motion. 
One is the Klein-Gordon equation for a conformally coupled massless scalar field and the other is
the Dirac equation for a massless fermion. Because of the absence of a
propagating scalar degree of freedom in the conformal limit, the scalar
field was set to zero by hand in \cite{Horava:2007ds}. 
The scalar is related to the ``tachyon'' in the effective 2D type 0A
theory and is only expected to completely
decouple in the strict $\alpha'\to\infty$ limit, which could be made
manifest  by setting $\phi=\varphi/(\alpha')^{1/4}$ in the following.

\subsection{Conformal gravity with matter} 

The theory of interest in this paper is a truncation  of this
higher spin theory  where we consider  just the spin-2 mode
and the  scalar and fermionic matter fields.  Actually,  the fermion will  not play any role in the following.
Recall that restricting the higher spin CS theory \eqref{HCSaction} to just the spin-2 mode one gets
conformal gravity in 2+1 dimensions \cite{Horne:1988jf},  a topological gravity theory governed by the
Chern-Simons action 
\eq{
                   S^{SO(3,2)}_{\rm CS}= {1\over 4} \int_{M} {\rm Tr}\left(A\wedge dA +{2\over 3} A\wedge
                   A\wedge A\right)
}
with the one-form $A$ taking values in the 3D conformal algebra $SO(3,2)$.
The resulting equation of motion says that the Cotton tensor is vanishing
\eq{
            C_{\mu\nu}=\epsilon_\mu{}^{\alpha\beta} \,\nabla_\alpha
            \Big(R_{\beta \nu}-{1\over 4} g_{\beta\nu}  R\Big) =0\,,
}
which is satisfied for a  conformally flat metric. Note that being
topological, this theory does not admit propagating degrees of
freedom.  
Moreover,  by being conformal it does not have a dimensionful
gravitational coupling like the Planck mass.

Now, one can couple this gravity theory to fermionic and bosonic
matter fields. Here we focus on a scalar field, whose  
action takes the following form 
\eq{
                  S_{\rm eff}=S^{SO(3,2)}_{\rm CS}+S_{\rm ferm}-\int d^3x &\sqrt{-g} \Big( {1\over 2} g^{\mu\nu}\partial_\mu
               \phi \partial_\nu \phi  \\
        &+ {\xi_3\over 2} R\, \phi^2+{\kappa\over 6} \phi^6 \Big)\,
}
with $\xi_3=1/8$ and 
where we also added a conformal, sixth order self-coupling of the field $\phi$
with a dimensionless coupling constant $\kappa$.  From now on
we suppress the contribution from the fermionic matter field.
One can show that this action is indeed invariant under a local
Weyl-transforma\-tion
\eq{
                    g_{\mu\nu}\to e^{2\Lambda(x)}  g_{\mu\nu}\,,\qquad
                    \phi\to e^{-\Lambda(x)/2} \phi\,.
}
It was argued in \cite{Nilsson:2015pua}
that also for the higher spin generalization, such an interaction term
can be consistently introduced.
Varying now $S_{\rm eff}$ with respect to the metric, one
arrives at the gravity equation of motion $C_{\mu\nu}-T^{\rm
  bos}_{\mu\nu}=0$,  that in detail takes the form 
\eq{
\label{eommetric}
     & C_{\mu\nu}-{\phi^2\over 8} \Big(R_{\mu\nu}-{1\over
          2}g_{\mu\nu} R\Big)  +{\kappa\over 6} \phi^6 g_{\mu\nu}\\[0.1cm]
   &- \Big(\partial_\mu
        \phi \partial_\nu\phi -{1\over 2} g_{\mu\nu} (\partial \phi)^2
        \Big)-{1\over 4}\Big( \phi\Box\phi +(\partial \phi)^2 \Big) g_{\mu\nu}\\
   &+{1\over 4}\Big(\phi \nabla_\mu \nabla_\nu\phi+\partial_\mu
    \phi\partial_\nu \phi\Big)=0\,.
}
It is remarkable that the Einstein tensor reappears in the equation of
motion, though not from varying the Chern-Simons gravity action but
from varying the action for the conformally coupled scalar.
However, the sign in front of the Einstein tensor is opposite
so that  effectively gravity acts repulsively. If $\phi$ carries a
non-vanishing vacuum expectation value $\phi_0$, then the prefactor
of the Einstein tensor can be thought of as an effective Planck scale
\eq{
\label{planckeff}
                       \widetilde M_{\rm pl}=-{\phi_0^2\over 8}\,.
} 
Taking the trace of \eqref{eommetric} one finds an expression that
vanishes if the equation of motion for the scalar field is obeyed
\eq{
\label{eomphi}
        \Box \phi - {1\over 8} R \phi -\kappa\phi^5 =0\,.
}
This is of course a consequence of the underlying Weyl symmetry.

\subsection{\textit{AdS} and \textit{dS} solutions}

Let us consider  potential $AdS_3$ and $dS_3$ solutions of radius $\ell$.
These are conformally flat and have  a constant Ricci scalar,  so the equation of
motion for $\phi$  admits the non-trivial solution
\eq{
\label{condphi}
\phi_0^4={3\over 4|\kappa|\ell^2}\,.
}

Here $AdS_3/dS_3$  has curvature $R=\mp 6/\ell^2$ so that one needs
$\kappa>0$  for $AdS_3$ and $\kappa<0$  for $dS_3$.
One can show that 
also the gravity equation of motion \eqref{eommetric} is satisfied.

Since $\ell$ sets a
length scale, this can be thought of as breaking the Weyl symmetry 
spontaneously, with the flat direction in the $(\phi,\ell)$ plane 
giving a Goldstone mode.  
Note that conformal gravity equipped with an additional scalar field with non-vanishing
vacuum expectation value $\phi_0\ne 0$ is nothing else than the well
known theory of  topologically massive gravity (TMG), whose action is usually written as
\eq{
          S_{\rm TMG}&=  {\widehat M_{\rm pl}\over 2}\int d^3 x \Big[ \sqrt{-g} \big(
          R-2\Lambda\big) \\&+{1\over 2\mu}\epsilon^{\mu\nu\rho}\Big(
          \Gamma^\alpha_{\mu\beta}\partial_\nu\Gamma^\beta_{\rho\alpha}
       +  {2\over 3} \Gamma^\alpha_{\mu\gamma} \Gamma^\gamma_{\nu\beta}
      \Gamma^\beta_{\rho\alpha}\Big)\Big]\,.
}
Note that the Weyl symmetry can  be used to eliminate fluctuations of the field $\phi$.
By comparison,  one finds a negative Planck scale $\widehat M_{\rm
  pl}=\widetilde M_{\rm pl}$ and 
\eq{
\label{TMGparam}
       \mu={\phi_0^2\over 8}={1\over 16\ell}\sqrt{3\over |\kappa|}\,,\qquad \Lambda=\mp {1\over \ell^2}\,.
}
This theory was first introduced in
\cite{Deser:1981wh}, where it was shown that  around a Minkowski
solution it has one massive propagating spin-2 degree of freedom 
of  mass $m_{(2)}=\mu$.
Thus,  as expected for a local symmetry,  the Goldstone mode does not survive as a massless mode.

\section{Swampland conjectures}

In this section we confront this effective almost topological theory with various
swampland conjectures, in particular those dealing with $AdS$ and
$dS$ solutions.

\subsection{\textit{AdS} swampland conjectures}

\begin{figure}[ht]
  \centering
  \includegraphics[width=7cm]{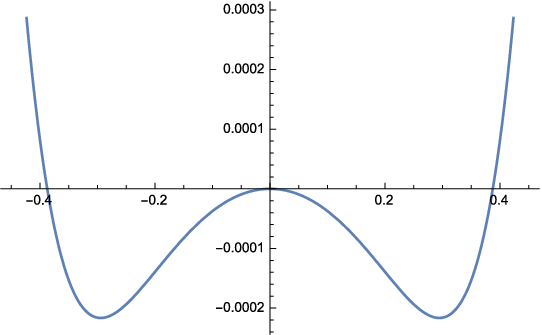}
\begin{picture}(0,0)
    \put(0,51){$\phi$}
      \put(-107,127){$V$}
    \end{picture}
  \caption{The potential $V(\phi)$ for $\kappa=1$ and $AdS$ length scale
    $\ell={10}$.}
  \label{fig:Veff}
\end{figure}
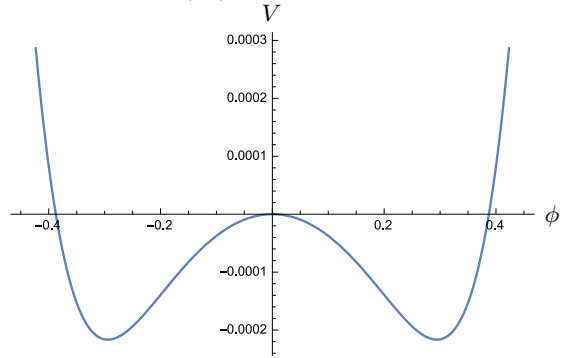

Thus, we consider the case $\kappa>0$ so that the action allows us to read off an effective
potential 
\eq{
\label{poteffi}
V(\phi)=-{3\over 8\ell^2}  \phi^2+{\kappa\over 6} \phi^6\,.
}
The form of this  potential is shown in figure \ref{fig:Veff},
where the non-trivial solution \eqref{condphi}  for $\phi_0$  is
evident.

It can readily be checked that the $AdS$ minimum satisfies the $AdS$/moduli scale
separation conjecture \cite{Gautason:2018gln} which  in our case says $m_\phi\,\ell\le
c$, for some $O(1)$ constant $c$. Just from $V(\phi)$ we find $m^2_\phi=3/\ell^2$.

Next, we take a look at 
the $AdS$ distance conjecture \cite{Lust:2019zwm}. 
From \eqref{TMGparam} we have for the cosmological constant $|\Lambda_{\rm
  AdS}|\sim \phi_0^4$. Therefore it vanishes for $\phi_0=0$ which,
opposed to the $AdS$ distance conjecture, is 
at finite distance in the moduli space. 

The $AdS$ distance conjecture  also says that
there should be  a tower of light states with scaling $m\sim
|\Lambda|^\alpha, \alpha>0$.
From the (refined) swampland distance conjecture \cite{Ooguri:2006in,Klaewer:2016kiy}, one expects a tower of states
with masses
\eq{
                        m=m_0 \,e^{-{\Delta\phi/ \sqrt{|\widetilde M_{\rm pl}}|}}\sim
                        \phi_0^2 \sim {1\over \ell} \sim \Lambda_{\rm AdS}^{1\over 2}
}
where we have used that the only mass-scale in the problem is $m_0\sim
\phi_0^2$ and the effective Planck scale \eqref{planckeff}. 

Since non-critical M-theory is defined in 3D,
there is no compactifcation and hence no tower of Kaluza-Klein modes
involved here. A natural candidate for this tower of states
are the infinitely many higher spin fields in the Chern-Simons action.
For the spin-2 field, the mass of the single massive degree of freedom
of TMG with a negative cosmological constant was computed in \cite{Carlip:2008jk,Carlip:2008eq}
\eq{ 
\label{massads}
m^2_{(2)}=\Big(\mu+{2\over \ell}\Big)^2-{1\over \ell^2}=
 {1\over \ell^2}(\mu\ell+3)(\mu\ell+1)
}
with $\mu\ell=\sqrt{3/(256\kappa)}$. 
This can be thought of as the $\mu\ell$-corrected mass of the field $\phi$ that
we read off from the effective scalar potential \eqref{poteffi} as
$m^2_\phi=3/\ell^2$. Moreover, the relation \eqref{massads} correctly
reproduces the mass of the graviton in the Minkowski limit $\ell\to\infty$.

Since we have $\mu\ell>0$,  this mass can never vanish. Thus, 
our case is different from the chiral model discussed in
\cite{Li:2008dq}, which in contrast to our case  had  a positive Planck mass.
Taking some results for higher spin topologically massive 
gravity \cite{Chen:2011vp,Bagchi:2011vr,Bagchi:2011td}
into account, a natural conjecture for the mass of the higher spin fields
would be
\eq{
m^2_{(s)}= {(s-1)\over
  \ell^2}\Big((s-1)\mu\ell+(s+1)\Big)\Big(\mu\ell+1\Big)\,,
}
which for $s\gg 1$ indeed gives the desired  scaling $m_{(s)}\sim
s\mu\sim s\phi_0^2$. Of course, a more thorough computation
in the framework of the full ${\cal W}_0$ higher spin theory is necessary to 
really confirm this relation.

Therefore,  consistent
with the finite distance in moduli space, there is only  a
polynomially and not  an exponentially
light tower of states. This is analogous to a similar
polynomial scaling  found for the finite distance conifold point in
\cite{Blumenhagen:2019qcg}. Similar to that  case, applying the emergence
proposal \cite{Heidenreich:2017sim,Grimm:2018ohb,Heidenreich:2018kpg} 
will lead us to a field dependent UV cut-off 
\eq{
\Lambda_{\rm UV}\sim \phi_0^2\sim {1\over \ell}
}
so that the number of light species is constant
and does not depend on the field $\phi_0$. This cut-off is also
consistent with the spin-2 swampland conjecture proposed in \cite{Klaewer:2018yxi}.

For 3D gravity with negative cosmological constant, it is known that there exist BTZ black-holes
which are locally isometric to $AdS_3$ space. Therefore, they will
also be solutions of TMG and also of the full equations of motion
\eqref{eommetric} and \eqref{eomphi}. However, with the wrong sign of the
Einstein-Hilbert term they will carry negative energy and therefore signal
a non-perturbative  instability of the $AdS_3$ background.
Let us remark that this is consistent with the proposed instability
\cite{Ooguri:2016pdq} of non-supersymmetric $AdS$-backgrounds.

To summarize, up to the finite distance in moduli space, this
almost topological higher spin theory has all the ingredients
to satisfy the $AdS$ swampland conjectures.

\subsection{\textit{dS} swampland conjecture}

Now let us choose $\kappa<0$ and  consider the
non-trivial $dS_3$  solution. Writing 
$\ell=1/H$ the action features a potential 
\eq{
\label{dsveff}
V(\phi)={3\over 8} H^2 \phi^2-{|\kappa|\over 6} \phi^6\,
}
which is just the one from figure \ref{fig:Veff} flipped at the $x$-axis.
Therefore, we have a $dS$ maximum and can check whether the (refined)
$dS$ swampland conjecture \cite{Obied:2018sgi,Garg:2018reu,Ooguri:2018wrx} 
in 3D is satisfied, i.e. in particular the relation
\eq{
\label{refdsswamp}
                     M_{\rm pl} |V''| > c'\,  V\,.
}
Computing both sides we find $V(\phi_0)={\kappa\over 3} \phi_0^6$ and 
$V''(\phi_0)=-4\kappa \phi_0^4=-3/\ell^2$. Using now the effective Planck scale \eqref{planckeff}
we get that both sides scale as $\phi_0^6$ and that \eqref{refdsswamp}
is satisfied if $c'<3/2$. The value of $c'$ is not known. One proposal
\cite{Blumenhagen:2020doa} is $c'=1/(D-1)(D-2)$ which for $D=3$ yields $c'=1/2$.

\subsection{Quantum break time}

It has been proposed in \cite{Dvali:2017eba,Dvali:2018fqu,Dvali:2018jhn} that $dS$ solutions in quantum gravity
experience the phenomenon of quantum breaking. This is argued to come from the
decoherence of the $dS$ bound state of graviton species. 

First we notice that since $dS_3$ is conformally flat, it is a solution to pure
conformal gravity and without any additional propagating modes present
it would be eternal and no quantum breaking would occur. However, 
after conformally  coupling it to a scalar field, one finds a
massive spin-2 degree of freedom so that a priori quantum breaking can occur.

Since the only scale in
the problem is set by $H$ or $\phi_0^{2}$, respectively,
just by dimensional  reasoning the quantum break time
should scale as $t_Q\sim H^{-1}$.  At first sight this is different
from the quantum break time $t_Q\sim {M_{\rm pl}/H^2}$ 
obtained in \cite{Dvali:2017eba} for a three-dimensional Einstein gravity theory. 
However, taking into account the effective Planck scale \eqref{planckeff} of
TMG, the two expressions indeed agree.

This finding is consistent with the proposal in \cite{Blumenhagen:2020doa}, namely that
in the high temperature regime of string theory, the quantum break
time should scale like  $t_Q\sim H^{-1}$ irrespective of the number of
dimensions. 
Due to the censorship of quantum breaking, there should be a
classical mechanism leading to a faster decay. From the form of the
potential in figure \ref{fig:Veff} it is clear that there exists a
tachyonic instability (in the $\phi$ direction) of mass $m^2= -3H^2$ leading indeed to a decay time
$t_{\rm dec}\sim H/|m|^2\sim H^{-1}$.

\subsection{Higuchi bound}

Finally, let us make a short comment on the Higuchi bound \cite{Higuchi:1986py,Lust:2019lmq}.
In particular, it says that in  a $d$-dimensional $dS$ background
there exists a lower bound for the mass of  an helicity $t$ mode of a massive  field of
spin $s$
\eq{
                   m_{(s,t)}^2\ge  H^2 (s-t-1)(s+t+d-4)\,.
}
Since in our case the sign of the Einstein-Hilbert term is reversed,
we expect that ghost freedom in $dS$ is guaranteed in the opposite
regime,  i.e. for the helicity $t=s$ mode in 3D we get
\eq{
\label{higushis}
                   m_{(s,s)}^2\le  -H^2 (2s-1)\,.
}
Next we need to know the mass of the spin-2 mode for TMG.
Redoing the computation from \cite{Carlip:2008jk,Carlip:2008eq}
for positive cosmological constant $\Lambda=\ell^{-2}=H^2$, we find
\eq{ 
\label{massds}
m^2_{(2)}=-\Big(\mu+{2\over \ell}\Big)^2+{1\over \ell^2}\,,
}
which correctly reproduces the mass of the tachyonic mode in the 
$\mu\ell\to 0$ limit. This calls for the higher spin generalization
\eq{ 
\label{massdsgen}
m^2_{(s)}=-\Big((s-1)\mu+{s\over \ell}\Big)^2+{1\over \ell^2}\le
(-s^2+1) H^2\,.
}
which for $s\ge 2$ indeed satisfies the (sign reversed) Higuchi bound \eqref{higushis}.

\section{Conclusions}

In this letter we have studied a well motivated (toy) model for the
high energy phase of string theory regarding its consistency with some
of the swampland conjectures.  We considered the truncation of the 
higher spin Chern-Simons theory to just the spin-2 mode equipped with
a conformal, self-interacting scalar field. For this model we
found $AdS$ and $dS$ vacua  that, up to some slight modifications, have a
good chance to satisfy the known swampland criteria.  Some steps
like the determination of the masses of the higher spin modes were left open.
It would also be interesting to consider similar models of this type,
as for instance the four-dimensional
Weyl-squared  action \cite{Stelle:1977ry,Ferrara:2018wqd,Ferrara:2018wlb}. 

The result of this letter can be interpreted
in two ways. First,  it demonstrates that there is a good chance for  the
existence of non-standard
effective gravity theories in the landscape of string theory.  Second, it
gives credence to the proposal that the high energy phase of string
theory does also satisfy the swampland criteria.

Turning the logic around and  using the swampland conjectures as a guide to the high
temperature phase, we conclude that a purely topological theory without
any propagating degrees of freedom might be too simple.

\section*{Acknowledgements}

We thank Dieter L\"ust  for useful comments. The work of RAG is supported in part by
the Deutsche Forschungsgemeinschaft under Germany’s Excellence Strategy EXC
2121 Quantum Universe 390833306.

\bibliography{references}
\bibliographystyle{utphys}

\end{document}